\documentclass{jaa}
\usepackage[utf8]{inputenc}
\usepackage{amssymb}
\usepackage{amsfonts}
\usepackage{amsmath}
\usepackage[T1]{fontenc}
\usepackage[british]{babel}
\usepackage{latexsym}
\usepackage{times}
\usepackage{etex}
\usepackage{booktabs}
\usepackage{balance}
\usepackage{float}
\usepackage{textcase}
\usepackage{txfonts}
\usepackage{caption}
\usepackage{color}   
\usepackage{epsfig}
\usepackage{graphics}
\usepackage{graphicx}
\usepackage{xmpmulti}
\usepackage{aas_macros} 
\usepackage{astrobib} 
\def\i{\item}

\newcommand{\bed}{\begin{displaymath}}
\newcommand{\eed}{\end{displaymath}}
\newcommand{\bei}{\begin{itemize}}
\newcommand{\eei}{\end{itemize}}
\newcommand{\bef}{\begin{figure}}
\newcommand{\eef}{\end{figure}}
\newcommand{\ben}{\begin{enumerate}}
\newcommand{\een}{\end{enumerate}}
\newcommand{\beq}{\begin{equation}}
\newcommand{\eeq}{\end{equation}}
\newcommand{\ber}{\begin{eqnarray}}
\newcommand{\eer}{\end{eqnarray}}
\newcommand{\bb}{\bf B}

\newcommand{\nb}{\bf \nabla}

\newcommand{\mdot}{\mbox{$\dot{M}$}}

\newcommand{\pdot}{\mbox{$\dot{P}$}}

\newcommand{\lsim}{\raisebox{-0.3ex}{\mbox{$\stackrel{<}{_\sim} \,$}}}
\newcommand{\gsim}{\raisebox{-0.3ex}{\mbox{$\stackrel{>}{_\sim} \,$}}}
\newcommand{\gcc}{\mbox{${\rm gm.} {\rm cm}^{-3}$}}

\newcommand{\msun}{\mbox{{\rm M}$_{\odot}$}}
\newcommand{\rsun}{\mbox{{\rm R}$_{\odot}$}}
\newcounter{attnctr} 

\begin{document}\sloppy

\title{The Enigma of GLEAM-X~J162759.5-523504.3}

\author{Sushan Konar\textsuperscript{*}}
\affilOne{NCRA-TIFR, Pune, 411007, India.}

\twocolumn[{

\maketitle

\corres{sushan.konar@gmail.com}


\begin{abstract}
It is  proposed that GLEAM-X~J162759.5-523504.3, the  newly discovered
radio  transient   with  an   unusually  long  spin-period   (P$_s$  =
1091.1690s), can  be identified  to be  a Radio  Magnetar which  has a
dipolar surface magnetic field of $2.5 \times 10^{16}$~G.  It is shown
that  - a)  it  is possible  to  anchor  such a  strong  field at  the
core-crust boundary  of a  neutron star,  and b)  the energy  of field
dissipation can  explain the observed  luminosity (radio \&  X-ray) of
this source.
\end{abstract}

 }]


\doinum{12.3456/s78910-011-012-3}
\artcitid{\#\#\#\#}
\volnum{000}
\year{0000}
\pgrange{1--}
\setcounter{page}{1}
\lp{1}

\section{Introduction}
Since the first (serendipitous) discovery of a neutron star as a radio
pulsar~\cite{hewis68},  some $\sim$3500  of  them are  now known  with
emissions   detected   across   almost  the   entire   electromagnetic
spectrum~\cite{manch05a}.  Despite  the wide variety  of observational
characteristics, these  are understood  to have  three basic  modes of
energy generation.   Accordingly, they can  be classified as -  a) the
rotation powered pulsars (RPP) where  rotational energy is lost due to
electro-magnetic braking; b) the accretion powered pulsars (APP) where
material accretion from a companion  gives rise to energetic radiation
and c)  the internal energy  powered (IEP) objects where  the emission
comes   from  certain   internal   reservoirs  of   energy  like   the
post-formation residual heat or energy stored in ultra-strong magnetic
fields~\cite{kaspi10,konar16c}.

Indeed, these  classes are neither  mutually exclusive, nor  are their
evolutionary connections unknown. In  fact, the connection between the
RPP and  APP has been  studied through the  decades and is  quite well
understood (see  \citeN{konar17e} for a  review).  On the  other hand,
detailed  theory  of  magneto-thermal evolution  and  interconnections
between different types  of isolated neutron stars  is being developed
in recent years~\cite{pons09,kaspi10,vigan13}.  The target objects for
this line  of investigation include  radio pulsars (RPSR), as  well as
IEP  class objects  like the  Magnetars, the  Central Compact  Objects
(CCO)  and  the X-ray  dim  Isolated  Neutron Stars  (XINS).   Already
evidence  for direct  evolutionary connections  between the  IEPs with
RPSRs   have    started   accumulating~\cite{kaspi17,jawor22,abhis22,chowh22}.

Barring the  APPs, the  neutron stars are  typically observed  to have
their  spin-periods ($P_s$) and  the  surface  magnetic fields ($B_s$)
within the following range -
\bei
   \i $P_s  \sim 1.3  \times 10^{-3}  - 23.5$~s,
   \i $B_s  \sim 10^{7}  - 10^{15}$~G, 
\eei
with the IEPs  predominantly clustering near the upper  bounds of both
the  parameters. However,  slow RPSRs,  with  periods as  large as  or
larger than  typical IEPs, have begun  to be detected in  recent years
with advanced observational capabilities. In  fact, two RPSRs with the
longest periods  (of all  known neutron stars)  - J1903+0433  ($P_s$ =
14.05s) and J0250+5854 ($P_s$ =  23.54s) - have been discovered within
the last five  years~\cite{han21,tan18}. Moreover, Magnetar-like X-ray
bursts have  now been  detected from a  number of  high-magnetic field
RPSRs  and some  of  the Magnetars  have also  been  observed to  emit
periodic radio pulses like RPSRs (these are called `Radio Magnetars').
The  latest object  to  join  these in-between  objects  is the  radio
emitting  J0901-4046 with  a  spin-period of  75.88s  and an  inferred
surface        dipolar        field       of        $1.3        \times
10^{14}$~G~\cite{caleb22}. Clearly, the boundary  between the RPPs and
the IEPs is getting blurred.

Quite naturally, the discovery of GLEAM-X~J162759.5-523504.3 - a radio
transient with  an unusually  long period of  $\simeq$ 1091s,  in the
archival  data  of  the  `Galactic and  Extra-galactic  All-sky  MWA  -
Extended'  survey has  generated  great interest~\cite{hurle22}.   Not
only is  the period more  than an order  of magnitude larger  than the
longest known before.  In  combination with other measured parameters,
it  also makes  the identification  of this  source (within  any known
neutron star class) extremely difficult.

Therefore, it is of importance  to understand the evolutionary history
of this  source, assuming `typical'  values for various  parameters at
birth. On the other hand, it is  also important to know if this source
is the  first representative  of a hitherto  unknown class  of neutron
stars and  if so how it  connects to other known  populations.  In the
current investigation,  we try to  look for  answers to some  of these
questions within the current paradigm of neutron star physics.

To begin  with, in  \S2 we  enumerate all  the measured  parameters of
GLEAM-X~J162759.5-523504.3 (referred  to  as  GLEAM-X hereafter)  and
discuss the  problems of  considering this source  as a  regular RPSR.
In \S3  we  review some  of the  hypotheses  that have  been
suggested                to                 explain                the
source~\cite{loeb22,katz22,ekcsi22,genca22,ronch22,tong22}.         It
appears  that fallback  accretion induced  slow-down of  a newly  born
neutron star holds maximum promise for explaining this source.  In \S4
we discuss the  complexities associated with such a  scenario.  In \S5
we offer a  simpler solution, in the form of  an ultra-strong magnetic
field anchored near  the core-crust boundary, to  the apparent mystery
of GLEAM-X.  Finally we conclude in \S6.

\section{Puzzling Parameters}
The  important observational  parameters  of GLEAM-X,  as reported  by
\citeN{hurle22}, are as follows -
\bei
\vspace{-0.30cm}
\i Basic  Parameters : 
\bei 
\i $P_s =  1091.1690 \pm 0.0005$~s, 
\i $\pdot_s \simeq 6 \times 10^{-10}$~s/s (best-fit value), 
%
%
\i brightness variation on timescales $\lsim  0.5$s; 
\eei
\i Radiation  Characteristics (Radio) :
\bei
\i $L_{\rm  rad} \lsim  4 \times 10^{31}$~erg.s$^{-1}$  (maximum pulse
flux),
%
%
\i $T_B \sim 10^{16}$~K,
\i linear polarisation  $\sim 88\%$, without any  variation with pulse
phase or time,
\i two `on' intervals of emission of  $\sim$30 days with a 26 day null
interval in-between;
\eei
\i Radiation Characteristics (X-Ray) : \\
$L_X \lsim 10^{32}$~erg.s$^{-1}$ obtained by Swift XRT;
\eei
where, $L_{\rm rad}$, $L_X$ and $T_B$ denote the radio luminosity, the
X-ray luminosity and the brightness temperature respectively.

A consideration of these parameters have  given rise to some amount of
puzzlement.  The main concerns are about  - a) the mechanism of energy
generation, and b)  the reason behind the  extremely long spin-period,
as discussed below.

{\bf \em  Energy Generation  :} The  regularity of  emission, combined
with  brightness variation  at  time-scales of  $\sim$0.5~light-second
(translates to a region of  size $1.5 \times 10^{10}$cm), implies that
the emission is  originating from a rotating  compact object. Assuming
the source to be rotation powered, we have -
\ber
\dot{E_{\rm R}}
&=& - I \, \omega_s \dot{\omega_s} = 4 \pi^2 I \, P_s^{-3} \pdot_s \\
\label{e_edot}
&\lsim& 2.0 \times 10^{28} \, \mbox{erg.s$^{-1}$}\,, \nonumber
\eer
where $E_{\rm R}$, $I$ and $\omega_{\rm s}$ are the rotational energy,
the moment  of inertia and  the spin  angular frequency of  a rotating
object. Here, $E_{\rm  R}$ is calculated assuming a  canonical mass of
1.4~\msun \, and  radius of 10$^6$cm for the neutron  star which gives
$I_{\rm  NS}  \sim 10^{45}$gm.cm$^2$  ($I  =  0.4MR^2$ for  a  uniform
density sphere, which  a neutron star effectively  is).  Clearly, this
falls  short of  the actual  radio  luminosity of  GLEAM-X by  several
orders of magnitude, indicating that the  source can not be powered by
rotational spin-down  alone. Because  of this,  it has  been suggested
that the source could be a Radio  Magnetar and powered by the decay of
its strong magnetic field~\cite{ronch22}.

{\bf  \em Slow  Rotation :}  The  nominal lifetime  estimated for  the
source, assuming a constant surface magnetic field, is -
\beq
P_s/2 \pdot_s \simeq 3 \times 10^4 \, \mbox{yr}\,.
\eeq
This is in  accordance with the expectation. Because,  the upper limit
of the  X-ray luminosity measured  suggests an age greater  than $\sim
10^5$yr~\cite{potek20}. However, the inherent  assumption here is that
the star slows down due to electro-magnetic braking alone, appropriate
for a  RPSR.  But such  a scenario runs  into trouble with  the energy
requirement, as  seen earlier.  On the  other hand, if it  indeed is a
Magnetar  then the  assumption of  a constant  magnetic field  becomes
problematic.

%

%
\section{Suggested Scenarios}
A  number   of  scenarios have   been  proposed  to   explain  the peculiarities  of GLEAM-X.  Here,  we describe them in brief.

{\bf Hot  Sub-dwarf (HSD)  -} \citeN{loeb22} hypothesised  that GLEAM-X
could be an  HSD, a stellar core evolving towards  a white dwarf phase
but not yet fully degenerate, pulsar.   Assuming a typical HSD mass of
0.5\msun \,  and radius  of 0.3\rsun \,
one obtains
$\dot{E_{\rm R}} \simeq 3 \times 10^{36}$ \, \mbox{erg.s$^{-1}$}\,,
which is more  than sufficient for the energy  requirement of GLEAM-X.
Moreover, a)  the characteristic age  and b) the rotation  periods are
not uncommon amongst  HSDs. However, the argument  entirely depends on
the HSD having a large  scale dipolar magnetic field of$\gsim 10^8$~G.
So far, no HSD has been  observed to harbour a magnetic field anywhere
near this strength.

{\bf White  Dwarf (WD) -}  In a somewhat similar  vein, \citeN{katz22}
has suggested that GLEAM-X could be a WD pulsar.  With I$_{\rm WD} \sim
10^{50}$~gm.cm$^2$ (for typical WD mass and radius) one obtains
$\dot{E_{\rm R}} \simeq 3 \times 10^{34}$ \, \mbox{erg.s$^{-1}$}\,,
sufficient to power GLEAM-X.  Many white dwarfs with $P_s$ as large as
a few hundred seconds are also known.  Importantly, the Lorentz factor
inferred from the  temporal substructure of GLEAM-X  pulses implying a
large  radius of  curvature (assuming  the  emission to  be caused  by
curvature  radiation)  is  more  in  conformity  with  a  white  dwarf
hypothesis rater  than a neutron  star one.  However, once  again, the
inferred  magnetic field  ($\gsim 10^{11}$~G)  turns out  to be  three
orders of  magnitude larger than the  largest field known to  exist in
white dwarfs~\cite{ferra20}.

{\bf  Precessing  Magnetar -}  A  number  of  Magnetars are  known  to
experience  spin precession,  with  precession periods  $\sim 10^{4  -
  5}$s, as  a result of the  deformation of the star  caused by strong
internal  toroidal  magnetic  fields~\cite{brait09}.   \citeN{ekcsi22}
consider the measured  period of GLEAM-X to be  this precession period
instead of the actual spin-period. It is not very clear why the slower
precession period would show up in  the data in exclusion of the (much
faster) spin-period itself.

{\bf  Fallback Disc  -} The  presence of  fallback disks  around young
neutron stars has been invoked to explain a large variety of phenomena
in recent years. When a neutron  star is born in a supernova explosion
some of the explosion ejecta may fail to escape and fall back onto the
neutron star.  If this  material possesses sufficient angular momentum
a stable disk  can form.  Even though the spin  evolution of a neutron
star  is  usually  determined  by  the  electro-magnetic  braking  and
gravitational wave radiation, it can be significantly affected through
the        star's        interaction         with        such        a
disk~\cite{chatt00,alpar01,perna14,li21}.   The existence  of fallback
disks has been observationally established  by the discovery of such a
disk around 4U 0142+61~\cite{wang06}.



\citeN{ronch22}  and  \citeN{genca22}  have explained  the  ultra-long
spin-period of GLEAM-X invoking an efficient propeller-phase slow-down
of the  star through its  interaction with  the fallback disc.   In an
accreting system, the  pressure of the infalling  material is balanced
by the stellar magnetic field at the magnetospheric radius (Alfv\'{e}n
radius  in a  system  with  spherical symmetry).   The  motion of  the
accreting material (charged plasma) is channeled by the magnetic field
towards the magnetic poles of the  star at distances smaller than this
magnetospheric  radius.  However,  if the  spin-frequency of  the star
happens to be  higher than the Keplerian angular velocity  of the disk
at the magnetospheric radius then  co-rotation of the infalling matter
results in extraction  of angular momentum, leading to  a slow-down of
the star. (We  shall discuss the details of  this `propeller-phase' in
the next section.)

Assuming an initial  spin-period of 10ms, an  initial crustal magnetic
field of 10$^{12 - 15}$G and an initial disk accretion rate of 10$^{19
  - 29}$gm.s$^{-1}$, \citeN{ronch22} have shown that it is possible to
slow  a neutron  star  down  to GLEAM-X  periods  through an  extended
propeller-phase.  However, the disk is assumed to become inactive once
the required spin-period is reached.   An abrupt drop in the accretion
rate has  been alluded to as  the reason for such  inactivation. It is
not clear why that is likely to happen.

The fallback model  developed by \citeN{genca22} differs  from that of
\citeN{ronch22}  in various  details.  In particular,  \citeN{genca22}
take the details of the thermal state of the disk into account.  Doing
this, they are able to show that the GLEAM-X period can be reached for
a specific range  of the disk parameters (related  to the temperature,
the kinematic  viscosity and  the irradiation efficiency).   But, here
too the  source is expected  to arrive  at the current  period shortly
before  the disk  is  inactivated. Thereafter,  the  evolution of  the
source  would  be  entirely  due to  electro-magnetic  torque  with  a
$\pdot_s \sim 10^{-18}$~s/s  for a  dipolar magnetic  field of  $\sim
10^{12}$G. However,  the currently  measured source parameters  do not
appear to agree with a purely electro-magnetic evolution.


%
\section{Problematic Propeller-Phase}
To  understand the  true effect  of the  disc-field interaction,  on a
neutron  star, detailed  modeling  is required  which  must take  many
parameters into consideration. However, it  is possible to extract 
some of the  important features using simple  physical principles.  To
this  end,  we use  the  following  values  for various  neutron  star
parameters hereafter --
$M_{\rm NS} = 1.4~\msun$,
$R_{\rm NS} = 10^6~{\rm cm}$,
and $\mdot_{\rm Edd} = 10^{-8}  \msun/{\rm yr} \simeq 5 \times 10^{17}
\mbox{gm.s$^{-1}$}$\,,
where  $\mdot_{\rm Edd}$  is the  Eddington  rate of  accretion for  a
neutron star with  the above-mentioned mass and radius.  We ignore the
detailed physical properties of the disk, as well as the nature of the
magnetic field  (beyond assuming  it to  be a  simple dipole)  in this
discussion.

Material accretion  onto a neutron  star is channelised by  its strong
magnetic     field    from     the     Alfv\'{e}n    radius,     given
by~\cite{pring72,ghosh79} -
\ber
R_A  &=&  (2G)^{-1/7}  M_{\rm   NS}^{-1/7}  R_{\rm  NS}^{12/7}  B_{\rm s}^{4/7} \mdot^{-2/7} \nonumber \\
    &=& 1.01 \times 10^{10} \, B_{15}^{4/7} \left(\frac{\mdot}{\mdot_{\rm Edd}}\right)^{-2/7} \mbox{cm}\,,
\label{e_ra}
\eer
where, $\mdot$ is the disk accretion  rate and $B_{15}$ is the dipolar
surface magnetic field of the neutron star in units of $10^{15}$G. 

Plasma flows in along magnetic  field lines, that are co-rotating with
the star,  from the Alfv\'{e}n  radius.  This  can happen only  if the
Alfv\'{e}n radius is less than the light-cylinder radius, $R_{\rm LC}$,
defined to  be the  radius at which  $R_{\rm LC}.\omega_s$  equals the
velocity of light. Here, $\omega_s~(= 2\pi/P_s)$ is the spin-frequency
of the  neutron star. The field  lines can not co-rotate  with the star
beyond this radius. Therefore we must have -
\beq
R_A \le R_{\rm LC} \,.
\label{e_rarlc}
\eeq
Using eq.\ref{e_ra} this translates into the following relation - 
\beq
B_{15}^{-2} \, \left(\frac{\mdot}{\mdot_{\rm Edd}}\right)
\ge 13.80 \, P^{-7/2}_s \,. 
\label{e_mdotl}
\eeq

Furthermore, if  the Keplerian frequency at  Alfv\'{e}n radius happens
to  be smaller  than the  spin-frequency of  the star,  then the  disk
material  must  be  spun  up  for co-rotation.   This  is  achieved  by
extracting angular  momentum from the  star, thereby slowing  it down.
Defining Keplerian radius, $R_K$, as the radius at which the Keplerian
frequency equals the stellar spin-frequency, we obtain -
\ber
%
\omega^2_s &=& \frac{2 G M_{\rm NS}}{R^3_{\rm K}} \nonumber \\
\mbox{or,} \; \; R_{\rm K}  &=& \left(\frac{2 G M_{\rm NS}}{\omega^2_s}\right)^{1/3} 
= 2.11 \times 10^8 P_s^{2/3} \, \mbox{cm}\,.
\label{e_rk}
\eer
At any point in  the disk that has a radial  distance (from the centre
of  the star)  larger than  $R_K$,  the Keplerian  frequency would  be
smaller  than the  stellar  spin-frequency. Therefore,  $R_A$ must  be
larger than $R_K$  for effective angular momentum  extraction from the
star. In  other words, for  the propeller phase  to be active  we must
have -
\beq
R_A \ge R_{\rm K} \,.
\label{e_rark}
\eeq
Using eq.\ref{e_ra} and eq.\ref{e_rk} this gives - 
\beq
B_{15}^{-2} \, \left(\frac{\mdot}{\mdot_{\rm Edd}}\right)
\le 7.59 \times 10^5 \, P_s^{-7/3} \,.   
\label{e_mdotu}
\eeq	
\begin{figure}

\vspace{-1.0cm}  
\begin{center}
\includegraphics[width=550pt]{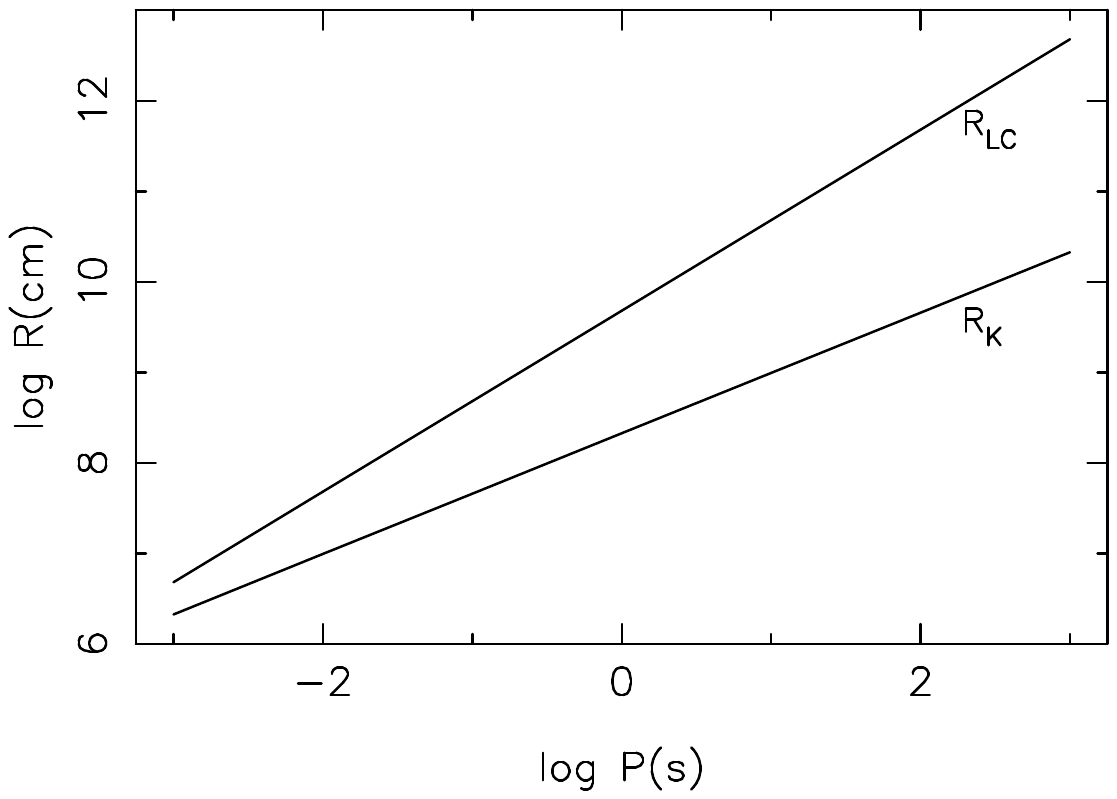}
\end{center}

\vspace{-7.0cm}
\caption{Variation of $R_{\rm LC}$ and $R_K$ with $P_s$, for a neutron
  star  of mass  1.4\msun \,  and radius  $10^6$cm. In  the propeller
  phase, $R_A$  must lie  within the  limits set  by $R_{\rm  LC}$ and
  $R_K$. See text for details.}
\label{f_rkrlc} 
\eef
Combining eq.\ref{e_rarlc} and eq.\ref{e_rark} we obtain the following
upper and lower limits for $R_A$ required for the propeller phase -
\beq
R_K \le R_A \le R_{\rm LC} \,.
\label{e_ralimit}
\eeq
Fig.[\ref{f_rkrlc}] shows the variation of $R_K$ and $R_{\rm LC}$ with
$P_s$ for  a neutron star  with our assumed  mass and radius.   In the
propeller phase, $R_A$ must always lie  within the limits set by these
two  lines. It  can  be seen  that the  difference  between $R_K$  and
$R_{\rm LC}$ increases with  increasing $P_s$.  At millisecond periods
$R_{\rm  LC}$  exceeds $R_K$  by  a  factor  of  2-3, but  the  factor
increases to 10  when $P_s$ reaches 10s. Clearly, $R_A$  is limited to
smaller intervals at shorter spin-periods compared to those for longer
spin-periods.

The full import of this is understood when we combine eq.\ref{e_mdotl}
and eq.\ref{e_mdotu} to find the bounds  on $B_s$ and \mdot \, for the
propeller phase. The condition is obtained as follows -
\beq
13.80 \, P^{-7/2}_s 
\le B_{15}^{-2} \left(\frac{\mdot}{\mdot_{\rm Edd}}\right) \le
7.59 \times 10^5 \, P_s^{-7/3} \,.   
\label{e_mdot_p}
\eeq	
In Fig.[\ref{f_mdot_p}] we have illustrated this for two values of the
dipolar  magnetic field.   The upper  shaded area  corresponds to  the
range of  \mdot \, for an  assumed field of $10^{16}$~G  and the lower
shaded  area corresponds  to  a  field value  of  $10^{12}$~G.  It  is
immediately seen  that, for propeller action, \mdot \, must  lie within
a specific range which changes rapidly with changing $P_s$.

It is,  of course, possible to  achieve a certain amount  of slow-down
with a  constant value  of \mdot  \, but it  depends crucially  on the
initial spin-period.  A very long  final period is achievable only for
moderately long initial periods. For example, final periods of $10^3$s
and  $10^4$s  (for  objects  like  GLEAM-X  and  1E  161348-5055)  are
achievable with constant \mdot \,, only from initial periods of 4s and
20s  respectively. Also,  starting from  a given  initial period,  the
maximum slow-down  is achieved  for a specific  value of  \mdot, which
changes significantly  with $B_s$.   To bring out  the same  change in
$P_s$, the \mdot \, required for  a $10^{16}$G field is about 8 orders
of magnitude higher than that required for a $10^{12}$G field.

\bef

\vspace{-1.0cm}  
\begin{center}
\includegraphics[width=575pt]{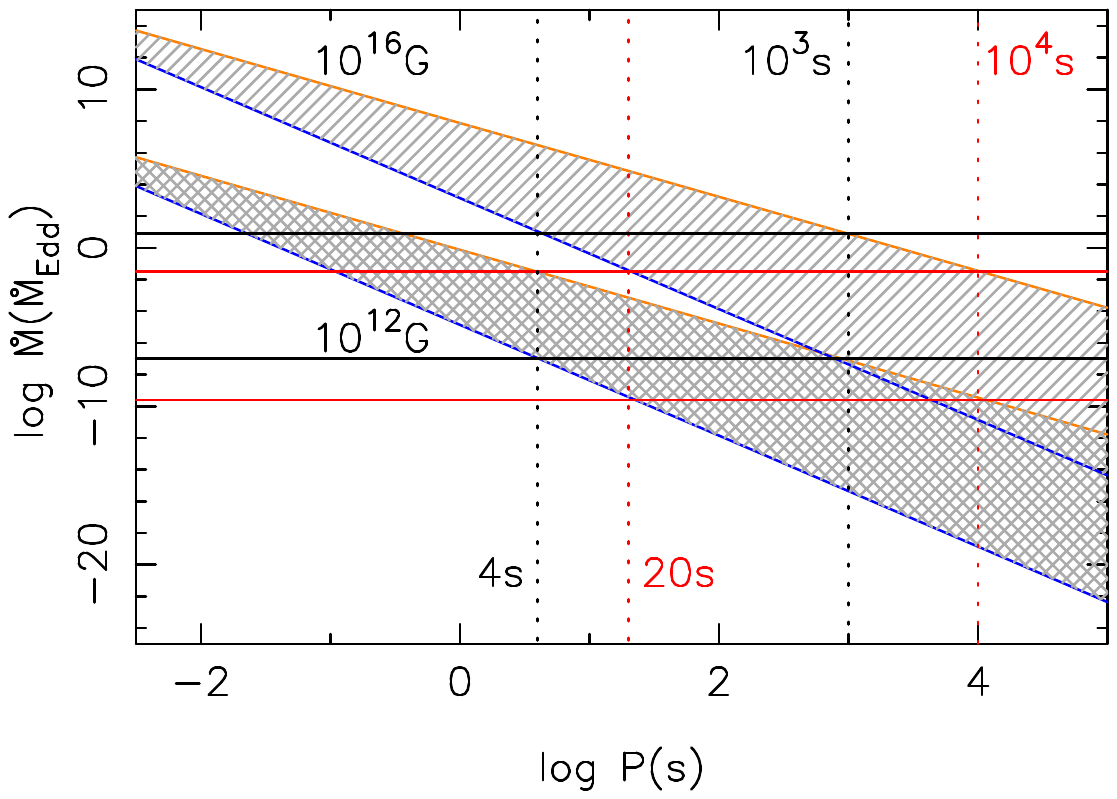}
\end{center}

\vspace{-7.0cm}  
\caption{$P_s$-dependent ranges  of \mdot  \, in the  propeller phase,
  for constant  surface magnetic fields.   The shaded areas  above and
  below  correspond  to  magnetic   field  values  of  $10^{16}$G  and
  $10^{12}$G  respectively.   The  vertical  broken  lines   in  black
  correspond  to $P_s$  values of  4s  and $10^3$s,  whereas those  in
  orange correspond to  $P_s$ values of 20s  and $10^4$s respectively.
  The \mdot \, values that can slow a neutron star down from an initial
  period to  a final  period are  indicated by  appropriate horizontal
  lines - black  lines for 4s to  $10^3$s and orange lines  for 20s to
  $10^4$s. }
\label{f_mdot_p} 
\eef

In short, spinning a neutron star  down to very long periods, starting
with  a  birth-period  in  milliseconds  (as  is  normally  expected),
requires  continuous fine-tuning  of  the  accretion rate.   Spin-down
results in  an increase in $R_K  \, (\propto P_s^{2/3})$. If  \mdot \,
does not decrease  appropriately then at one point  $R_K$ would become
larger  than $R_A$.   When that  happens,  material would  be able  to
accrete  onto the  polar regions  of the  star bringing  in additional
angular momentum.   Instead of propeller induced  spin-down this would
result in accretion-induced spin-up of the star. On the other hand, if
\mdot \, decreases  very rapidly then $R_A  \, (\propto \mdot^{-2/7})$
may  become larger  than  $R_{\rm LC}$  and the  disk  would then  get
disconnected from the star.

In summary, for a constant \mdot  \, system, the propeller phase would
be short-lived and the final period  would be determined by the period
at which the  system enters the propeller  phase.  Ultra-long periods
would only  be attained if  the neutron  star has already  been slowed
down significantly  (by electro-magnetic braking etc.).   On the other
hand, if a system goes through  an extended period of decreasing \mdot
\,, it may  oscillate between a propeller and an  accretion phase or a
propeller and a non-contact phase (unless the rate of decrease of \mdot
\, exactly equals  the value required for maintaining the  system in a
continuous  propeller  phase).   However, even  if  such  oscillations
finally result in slowing a neutron  star down to a period of $10^3$s,
it may take a much longer time than the estimated age of GLEAM-X.

\section{A Mighty Magnetar?}
As  has  been  noted  in  the discovery  paper,  the  extremely  large
brightness temperature  of GLEAM-X  implies coherent emission  and the
high percentage  of linear polarisation indicates  an ordered magnetic
field. Moreover,  the nature of  the pulse  profiles is thought  to be
suggestive of  this source being  a Radio Magnetar, even  though later
work by \citeN{erkut22} has shown  that the spin-down power of GLEAM-X
can support its observed radio luminosity like a regular RPP.

It is  therefore important to get  a measure of the  magnetic field of
this source.   Despite the use of  the best-fit value of  \pdot$_s$ to
estimate the spin-down energy, most  authors have considered it simply
to be an upper limit.  However, the exact value of \pdot$_s$ itself is
of significance.

If  one  assumes  the  spin-down  to  be  purely  electromagnetic  (no
observational  indication  for  ongoing accretion)  then  the  surface
dipolar field can be estimated to be~\cite{lorim04} -
\beq
B_{\rm s}
= \left(\frac{3  c^3}{8 \pi^2}  \frac{I_{\rm NS}}{R_{\rm  NS}^6 \sin^2
  \alpha} \, P_s \dot{P}_s \right)^{1/2} 
%
\simeq 2.5 \times 10^{16} \mbox{G}\,,
\label{e_dipol}
\eeq
obtained  assuming  $\langle  sin^2  \alpha  \rangle  \sim  1$,  where
$\alpha$ is  the inclination angle  (between the axes of  rotation and
dipolar magnetic  field).  This is  an order of magnitude  larger than
the  largest magnetic  field known  for a  neutron star  ($2.06 \times
10^{15}$~G for J1808-2024,  a Magnetar) till now.

Fig.[\ref{f_bp}] shows  all known  RPSRs and  IEPs for  which magnetic
field estimates are available.  Two death-lines, most relevant for the
known   neutron   star  population,   have   also   been  shown   (see
\citeN{konar19e}  for a  detailed discussion  on the  many death-lines
extant  in the  literature  and their  significance).   Both of  these
death-lines are developed by \citeN{chen93a} and are obtained assuming
pair  productions ($\gamma  + B  \rightarrow  e^- +  e^+$, $\gamma$  -
photon, $e^{-/+}$  - electron/positron, B -  magnetic field), required
for pulsar  emission, to occur predominantly  near the polar cap  of a
neutron star~\cite{ruder75}.   Assuming - a)  the surface field  to be
dipolar, and  b) the radius of  curvature of the magnetic  field to be
approximately  equal  to  the  stellar  radius,  the  death-lines  are
obtained as follows -
\ber
\mbox{\bf DL-1} &\mbox{\bf :}& 4 \log B_s - 6.5 \log P_s = 45.7 \,, \\
\label{eq-dl1}
\mbox{\bf DL-2} &\mbox{\bf :}& 4 \log B_s - 6 \log P_s = 43.8 \,;
\label{eq-dl2}
\eer
where $P_s$ is in  seconds and $B_s$ is in Gauss;  with {\bf DL-1} and
{\bf DL-2} corresponding  to the cases of very curved  field lines and
extremely twisted field lines respectively.

\begin{figure*}[h]

\vspace{-1.0cm}
\includegraphics[width=525pt]{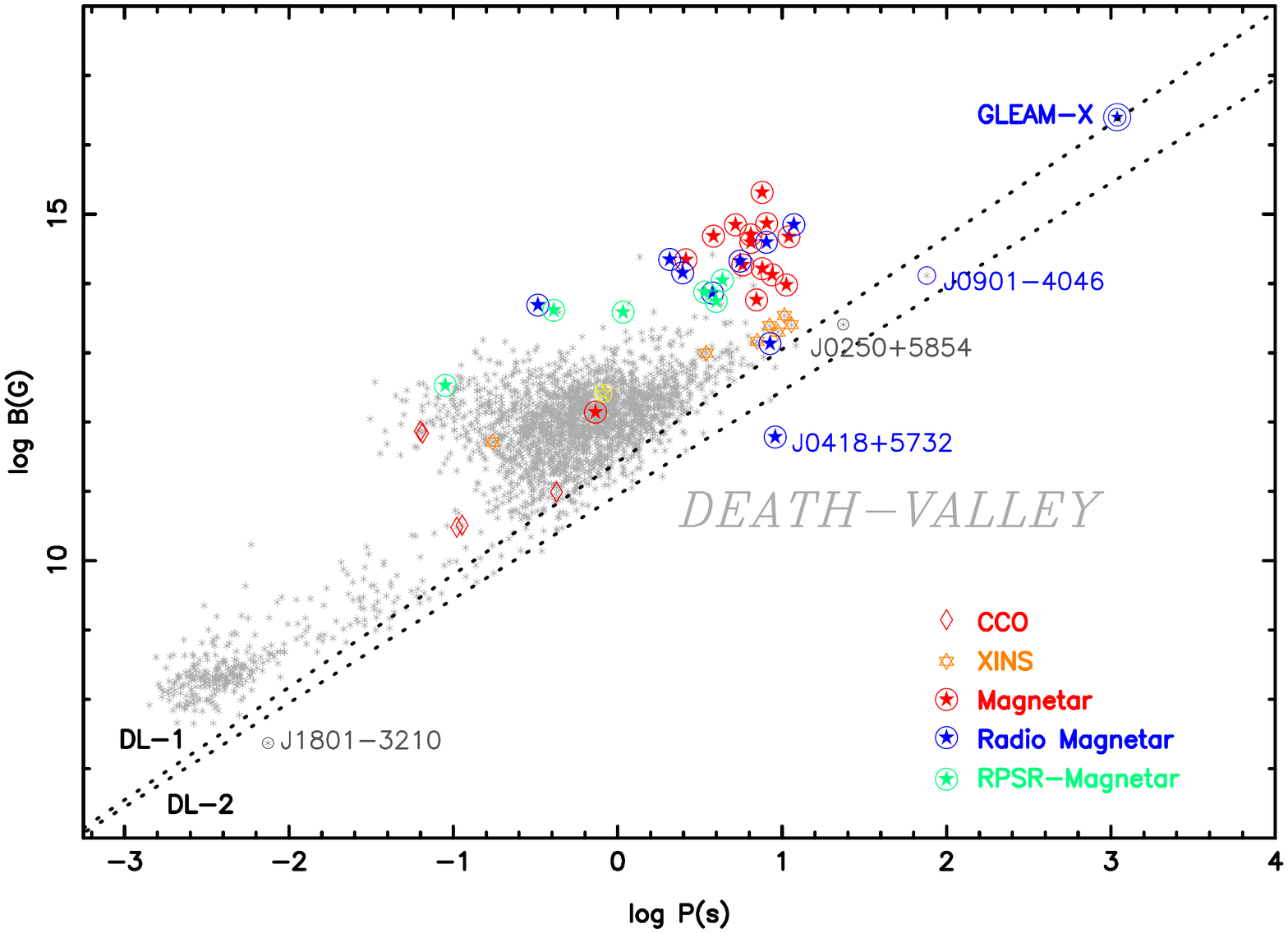}

\vspace{-0.25cm}
\caption{Distributions of RPSRs (grey star)  and IEPs, for which $B_s$
  estimates exist, in the $P_s$--$B_s$ plane.  Different categories of
  IEPs (CCOs,  XINS, Magnetars) are shown  separately, including three
  types of Magnetars - a) regular Magnetars, b) Radio Magnetars and c)
  RPSRs  with Magnetar  like bursts  (RPSR-Magnetar).  {\bf  DL-1} and
  {\bf DL-2}  are two death-lines  discussed in the  text.  J1801-3210
  (millisecond RPSR) and J0418+5732 (Radio Magnetar) are two prominent
  objects on the right of {\bf DL-2}.  J0250+5854 is the slowest known
  RPSR ($P_s$ =  23.5s) and J0901-4046 is the radio  neutron star with
  $P_s$ = 75.88s. \\
  {\bf Data Source :} a) {\tt https://www.atnf.csiro.au/research/pulsar/psrcat/}, \\
  b) {\tt http://www.physics.mcgill.ca/$\sim$pulsar/magnetar/main.html}, \\
 c) Gotthelf, Halpern, \& Alford 2013; De Luca 2017.} 
\label{f_bp} 
\end{figure*}
\nocite{gotth13,deluc17}

It can  be seen that  the majority of the  neutron stars are  bound by
{\bf  DL-1} while  the rest  are bound  by {\bf  DL-2}, including  the
slowest pulsar J0250+5854.  Only a  handful are seen beyond {\bf DL-2}
(in particular, the Radio Magnetar J0418+5732 and the millisecond RPSR
J1801-3210), and none of these  are deep into the pulsar death-valley.
It should be noted that these  death-lines are not meaningful for APPs
and those  have therefore been excluded  from this figure. It  is well
known  that  the X-ray  pulsars  in  HMXBs  function far  beyond  {\bf
  DL-2}~\cite{konar17e}.

Now, these death-lines correspond to  the cessation of pulsar emission
due to scarcity  of pair-production. Of course, the  emission from the
IEPs are not  necessarily due to such a process  (for example, the XINS
emission is understood to be caused by the residual heat retained from
the  supernova explosion).  The  fact that  all non-accreting  neutron
stars are  more or less bound  by these death-lines, with  most of the
IEPs located away  from them, simply mean that the  RPSRs tend to have
the  smallest surface  magnetic fields  for  a given  value of  $P_s$.
Therefore, a Magnetar  candidate is ordinarily expected  to be located
well above  {\bf DL-2}.   Using the surface  field estimated  from the
best-fit \pdot$_s$, we find that GLEAM-X is located very close to {\bf
  DL-1}, comfortably  above {\bf DL-2}.  Moreover,  a definite nulling
phase  has been  detected in  the radio  emission of  GLEAM-X and  the
nulling  pulsars  are  known  to  be  found  predominantly  near  {\bf
  DL-1}~\cite{konar19e}.   Considering these,  the  estimate of  $B_s$
(and therefore the best-fit value of  \pdot$_s$) does not appear to be
completely unreasonable. 

We are therefore confronted with two questions - {\bf a)} whether such
a strong field can be supported by a neutron star, and, if so {\bf b)}
whether it is possible for GLEAM-X to function as a Radio Magnetar.

{\bf (a).}  The  maximum stable field anchorable in a  neutron star is
$\sim  10^{18}$~G~\cite{carda01}, if  the currents  supporting such  a
field is located deep within the  core region.  This magnetic field is
likely  to be  in  the  form of  Abrikosov  fluxoids,  not subject  to
dissipative       processes,        in       the       superconducting
core~\cite{shapi83}. Therefore, it is possible for GLEAM-X to function
as an  RPSR even though its  inferred surface field would  then be more
than two orders of magnitude higher than the highest known for an RPSR
($9.36\times 10^{13}$~G for J1847-0130).

However, a Radio  Magnetar is necessarily powered by the  decay of its
ultra-strong  magnetic field. Therefore,  we must  consider a  crustal
field,  for  which  dissipation  is possible  because  of  the  finite
transport             coefficients              of             crustal
material~\cite{pons07b,pons09,konar17e,pons19}.

The melting temperature ranges from $\sim 10^8 - 10^{10}$~K across the
crust.  So the  crust is  likely to  be in  a crystalline  state from
almost immediately after the neutron  star is formed.  This crystal is
understood  to be  composed of  unscreened  nuclei arranged  in a  bcc
`metallic' lattice with inter-nuclear spacing (varying from $10^{-9} -
10^{-12}$~cm in the density range  of $10^8 - 10^{14}$~\gcc) exceeding
the nuclear size  by several orders of magnitude, and  therefore to be
`Coulombic' in nature.

For  the   crust  to   anchor  a  magnetic   field  in   a  stationary
configuration, the anisotropic field pressure  must be balanced by the
non-axisymmetric stresses  generated in the crustal  solid.  Therefore
the strongest magnetic field, $B_{\rm  max}$, that can be supported by
the solid crust is -
\ber
P(B_{\rm max}) &\leq& \sigma_{\rm max}^{\rm shear} \nonumber \\
\mbox{or,} \; \; B_{\rm max} &\leq& \sqrt{8 \pi \sigma_{\rm max}^{\rm shear}} \,,
\eer
where $P(B)$ is  the magnetic pressure.  The `yield'  or shear stress,
$\sigma^{\rm shear}$, of the crustal solid is given by~\cite{chugu10} -
\beq
\sigma^{\rm shear} = \left(0.0195 - \frac{1.27}{\Gamma - 71}\right) \frac{n(Ze)^2}{a},
\label{e_sigma}
\eeq
where, $n$,  $a$, $Z$  and $\Gamma$  are the  ion number  density, the
inter-ionic  distance  (lattice spacing),  the  atomic  number of  the
dominant ionic species  (at a given density) and  the Coulomb coupling
parameter  ($\gsim 250$).   Clearly, the  maximum value  of the  shear
stress, $\sigma_{\rm  max}^{\rm shear}$,  is obtained for  the maximum
density at the  bottom of the crust ($\gsim  10^{14}$~\gcc), where the
dominant ionic species is that of $Z = 20$~\cite{onsi08}. Accordingly,
%
\ber
\sigma_{\rm max}^{\rm shear} &\simeq& 1.2 \times 10^{32}~{\rm dyne.cm}^{-2}\,, \nonumber \\
%
\mbox{or,} \; \; B_{\rm max} &\leq& 5 \times 10^{16}~{\rm G}\,. 
\eer
Therefore,  it is  possible for  GLEAM-X to  have a  crustal field  of
magnitude $2.5 \times  10^{16}$~G, if it is anchored at  the bottom of
the crust.

It  must be  noted  that  eq.\ref{e_sigma} has  been  obtained by  the
authors for  $^{56}$Fe matter  at a  density of  $10^9$~\gcc.  Earlier
calculations of $\sigma^{\rm shear}$  by \citeN{stroh91}, using strain
angle  estimates of  \citeN{smolu70b},  gave much  smaller values  for
$\sigma^{\rm shear}$.   However, \citeN{horow09b}  and \citeN{chugu10}
have established  that the crust is  expected to be much  stronger and
have much  higher values of  $\sigma^{\rm shear}$.  Therefore,  it may
not   be  entirely   unreasonable  to   extrapolate  the   results  of
\citeN{chugu10} to higher densities.

Incidentally, for a self-gravitating object  (like a neutron star) any
interior magnetic  field is  also subject to  the constraint  that the
field pressure must always be  smaller than the gravitational pressure
for the structural stability of the object~\cite{nitya14,nitya15a}. As
things  stand  today, it  is  possible  for any  one  of  a number  of
equations of state  to match the observed mass-radius  relation of the
known neutron  stars. However,  all of these  equations of  state have
pressures    upwards    of    10$^{33}$dyne/cm$^2$   for    $\rho    >
10^{14}$~\gcc~\cite{chame08}.  Consequently, a  magnetic field of $2.5
\times 10^{16}$~G is comfortably accommodated.

In  recent years,  there have  been predictions  of a  `nuclear pasta'
phase  in  the boundary  layer  between  the  crust  and the  core  at
densities  $\gsim  10^{14}$~\gcc.   Such  a  layer,  though  thin,  is
expected  to  have  much   higher  resistivity  and  lower  structural
strength,    compared    to  a crystalline   solid    at    equivalent
densities~\cite{horow15}. Therefore, in presence  of such pasta phase,
the strongest  field would  have to  be anchored  at a  somewhat lower
density where the crust behaves like a `Coulombic' solid.

{\bf (b).}  Assuming that it is  possible to anchor a  $\gsim 10^{16}$~G
field in the crust  as discussed above, it is necessary  to see if the
energy  of field  decay can  power GLEAM-X.  The evolution  of such  a
crustal field, in  absence of any material motion, is  governed by the
following equation~\cite{jacks75} -
\beq
\frac{\partial \bb}{\partial t}
=  - \frac{c^2}{4\pi}  \nb  \times  \left(\frac{1}{\sigma(\rho, T)} \nb  \times
\bb\right) \,,
\eeq
where, $\sigma(\rho, T)$  is the electrical  conductivity which depends  on 
density ($\rho$) and temperature ($T$) of the crustal material.  If the currents
supporting  the  field  has  a  small  radial  extent  such  that  the
conductivity  can be  taken to  be constant,  the time-scale  of field
evolution through Ohmic dissipation becomes -
\beq
\tau_{\rm Ohmic}
=  \frac{4\pi L^2 \sigma}{c^2} \,, 
%
\eeq
where $L$ is the radial extent of the region of current concentration.
The density  at the bottom of  crust is $\gsim 10^{14}$~\gcc,  and the
temperature at  that depth can  be assumed to  be $\sim 10^7$~K  for a
neutron star of age $\lsim 10^5$~yr. For these parameters, $\sigma$ is
about $10^{28}$s$^{-1}$ for a pure  crust, and $10^{25}$s$^{-1}$ for a
crust with 5\%  impurity. If, on the other hand,  the bottom layers of
the crust  remain at a much  higher temperature of $\sim  10^8$~K, the
conductivity of the pure crust drops to $\sim 10^{25}$s$^{-1}$ and the
effect  of impurity  can be  ignored  at such  high temperatures  (See
\citeN{chame08} for a detailed discussion on crustal micro-physics.).

Therefore, if we assume the radial extent of the current concentration
to be $\sim$10\% of the crust,  i.e, about 100m, the Ohmic dissipation
time-scale turns out to be $\sim 10^6 - 10^9$~yr. Then, the fractional
change in the field strength over the source life-time would be -
\beq
\frac{\delta B}{B}
= \frac{\tau_{\rm GLEAM}}{\tau_{\rm Ohmic}}
= 10^{-4} - 10^{-1} \,,
\eeq
where $\tau_{\rm  GLEAM}$ is the  estimated spin-down age  of GLEAM-X.
In other words, the  drop in the field strength would  be 10\%, at the
most.  So  it can  be considered  to be  effectively constant  for all
practical purposes.

However, even for such minimal change  in the field strength the total
energy of  dissipation could be  quite substantial.  The  total energy
stored in the field can be estimated as follows -
\beq
E_{\rm B} = \frac{B^2}{8 \pi} 4 \pi R^2 L
= \frac{1}{2} B^2 R^2 L 
\sim 10^{52}~\mbox{erg} \,,
\eeq
where $R$ is the radius of the star. Therefore, the average rate of field
dissipation energy available to the star, over its estimated lifetime, is
given by -
\beq
\frac{\delta E_{\rm B}}{\tau_{\rm GLEAM}}
\simeq \frac{\delta B \, B R^2 L}{\tau_{\rm GLEAM}} 
%
\sim 10^{35} - 10^{38} \mbox{erg.s$^{-1}$} \,.
\eeq
This is far in excess of the maximum observed radio luminosity ($\gsim
10^{31}$~erg.s$^{-1}$)      or       X-ray      luminosity      ($\sim
10^{32}$~erg.s$^{-1}$)  of GLEAM-X,  as  is typically  expected for  a
Radio Magnetar. Therefore, it is entirely possible for GLEAM-X to be a
Radio Magnetar,  albeit with an  ultra-strong magnetic field  of $\sim
2.5 \times 10^{16}$~G.

\section{Conclusions}
The newly  discovered radio  transient GLEAM-X  J162759.5-523504.3 has
given rise  to a number  of conjectures about its  nature. Uncertainty
regarding the true  nature of this source arises  primarily because of
its ultra-long  period and  the inadequacy  of the  inferred spin-down
power to explain its current luminosity.  Several authors have invoked
`fallback accretion induced slow-down' of a newly born neutron star to
successfully explain the ultra-long period of this source. However, we
find that spinning  a neutron star down to very  long periods requires
continuous fine-tuning of the accretion rate. Otherwise, final periods
of $10^3$s are only achievable from initial periods of 4s or larger if
the accretion rate remains constant.

In this work, we have considered a simpler solution to the problem, by
taking the  best-fit value  of \pdot$_s$  as the  true measure  of the
spin-down rate of the source.  Assuming a dipolar spin-down this gives
a  magnetic field  of $2.5  \times 10^{16}$~G,  an order  of magnitude
larger than the strongest field known in any neutron star.

However, we are able to demonstrate that -
\ben
\i the location  of GLEAM-X, in the $P_s-B_s$ plane,  is in conformity
with other neutron stars of the RPP as well as IEP class,
\i it is possible  to anchor such a strong field at  the bottom of the
crust of a neutron star,
\i  the energy  of  field  dissipation is  sufficient  to explain  the
observed luminosity (radio \& X-ray) of GLEAM-X.
\een
From this,  we draw  the conclusion  that GLEAM-X  is most  probably a
Radio Magnetar with an ultra-strong magnetic field.

It is to be noted that whether  a neutron star is powered by spin-down
(RPP) or by  the decay of its magnetic field  (Magnetar), the magnetic
field  plays a  crucial role  in  its spin-down.   In particular,  the
stronger  the  field, the  faster  would  be the  expected  spin-down.
Because  of this,  all the  slow neutron  stars, discovered  recently,
possess very strong magnetic fields.

For  example,  if  we  consider  $10^{14}  -  10^{15}$~G  fields,  the
spin-period attained (starting from an initial $P_s  \sim$~ms at birth)
at an age  equivalent to that of  the GLEAM-X would be  $\sim 50$s and
$\sim 200$s  respectively ($P_s$ scales  as $B$).  On the  other hand,
both $L_{\rm rad}$ (obtained from  spin-down) and $L_X$ (obtained from
field  decay) would  be down  by  factors of  $10^{-5}$ and  $10^{-3}$
respectively as they scale as $B^2$.

Quite interestingly, this scaling relation works out very well for the
newly  discovered  J0901-4046.  Its  surface  dipolar  field  and  the
characteristic age  have been estimated  to be $1.3  \times 10^{14}$~G
and 5.3~Myr respectively.  When these  values are used to estimate its
current spin-period  we obtain $\sim  70$~s - almost  exactly matching
the measured period of 75.88s!

\section{Acknowledgment}
The idea for this work arose out  of a couple of lively discussions on
GLEAM-X  J162759.5-523504.3  during  our weekly  COD  (Compact  Object
Discussion)  where a  group of  Indian neutron  star enthusiasts  meet
(virtually) to talk  about matters of interest.

\end{document}